\documentclass[runningheads]{llncs}
\usepackage[T1]{fontenc}
\usepackage{marvosym}
\usepackage{textcomp}
\usepackage{graphicx}
\begin{document}
\title{A Map of Exploring Human Interaction patterns with LLM: Insights into Collaboration and Creativity}
\titlerunning{A Map of Exploring Human Interaction patterns with LLM} % 页眉中的缩短标题
\author{Jiayang Li\thanks{Both authors contributed equally to this research.} (\Letter) \and
        Jiale Li\protect\footnotemark[1] \and Yunsheng Su} 
\authorrunning{J. Li et al.}
\institute{College of Design and Innovation, Tongji University, ShangHai, China
Jiayanglee0506@gmail.com}
\maketitle              % typeset the header of the contribution

\begin{abstract}
The outstanding performance capabilities of large language model have driven the evolution of current AI system interaction patterns. This has led to considerable discussion within the Human-AI Interaction (HAII) community. Numerous studies explore this interaction from technical, design, and empirical perspectives.
However, the majority of current literature reviews concentrate on interactions across the wider spectrum of AI, with limited attention given to the specific realm of interaction with LLM.
We searched for articles on human interaction with LLM, selecting 110 relevant publications meeting consensus definition of Human-AI interaction. Subsequently, we developed a comprehensive Mapping Procedure, structured in five distinct stages, to systematically analyze and categorize the collected publications. Applying this methodical approach, we meticulously mapped the chosen studies, culminating in a detailed and insightful representation of the research landscape.
Overall, our review presents an novel approach, introducing a distinctive mapping method, specifically tailored to evaluate human-LLM interaction patterns. We conducted a comprehensive analysis of the current research in related fields, employing clustering techniques for categorization, which enabled us to clearly delineate the status and challenges prevalent in each identified area.

\end{abstract}
%
%
%
%INTRO
\section{Introduction}
large language model(LLM) have significantly enhanced AI systems, particularly in generation, general intelligence, continous learning, and intention understanding\cite{chang_survey_2024}. Over the past two years, there has been a rapid increase in the number of studies and novel application designs revolving around LLM-based systems\cite{zhao_survey_2023}. Technical issues that previously plagued researchers have been effectively optimized or resolved with the advent of large language model\cite{min_recent_2021}.\\
\indent Amid rapid AI development, human-AI interaction is also evolving. Since a multitude of LLM-based solutions and system designs have emerged, expanding existing HAII patterns\cite{hadi_survey_2023}. However, comprehensive research on human interaction with LLM-based AI systems is lacking, as most HAII reviews focus broadly and overlook specific LLM applications.\\
\indent In this paper, we introduce a 5 stages mapping pocedure which includes defining the perspective, identifying dimensions, associating relevant concepts within dimensions, establishing evaluation criteria based on concepts, scoring and clustering.\\
\indent Following the pocedure, we selected 106 out of 1398 papers based on selection criteria that align with the general definition of Human-AI Interaction(HAII). We adopt the perspectives of collaboration and creativity, and establish dimensions of Human-AI Interaction and Implementation-Creation. We then gather concepts pertinent to these dimensions to formulate our evaluation criteria. Finally, we used manual scoring and machine clustering for mapping, leading to an Overview and Map of Human-LLM Interaction Research.\\
\indent Our mapping shows 4 clusters of human-LLM interaction: Processing tool, Analysis assistant, Creative companion and Processing agent. Overall, key contributions of this article include a new mapping procedure for human-AI interaction and a comprehensive map guided by this procedure. We anticipate this map will outline the field's emerging areas, aiding in the evaluation of current LLM systems and identifying future challenges and directions.
%RELATED WORK
\section{Related work}
\subsection{The Undergoing Change in HAII driven by large language model}
The advent of LLM and related technologies has led to rapid advancements in AI capabilities, particularly in the areas of general knowledge, intent understanding, and creative abilities. This has significantly altered the patterns and processes by which people interact with AI systems built upon LLM.
\subsubsection{General Intelligence — The capability to process complex tasks and the ability for continual learning.}
Early artificial intelligence mainly relied on machine learning algorithms such as statistical models\cite{muggleton_alan_2014}, with a general logic of constructing a model through learning from existing data to then infer about new data or samples\cite{mahesh_machine_2019}. These models, being trained on specific datasets, are typically constrained to the domain or task relevant to that data, acquiring new knowledge or learning new tasks depends on new training data and the retraining of the model\cite{niu_decade_2020}. Within this technical context, AI acts merely as a static tool\cite{smith_building_2017}, performing specific low-level tasks such as prediction and classification based on fixed, standardized input data, and then returning the analysis results to the user or to other modules within the system\cite{pang_thumbs_2002}\cite{li_survey_2022}\cite{noauthor_full_nodate}\cite{noauthor_automatic_nodate}.\\
\indent Unlike previous task-specific models and pre-trained models equipped with domain intelligence\cite{zhao_survey_2023}, LLM demonstrate advanced capabilities in various cognitive tasks due to their extensive training on diverse datasets and large parameter sets\cite{bubeck_sparks_2023}. LLM can perform various low-level tasks in a generative manner, invoking different external tools\cite{qin_tool_2023} or knowledge\cite{noauthor_210408164_nodate} as needed. In this context, LLM can autonomously manage certain aspects of complex tasks\cite{zhou_least--most_2023}, strategically utilizing their diverse abilities, and orchestrating these capabilities to execute high-level tasks\cite{noauthor_230317760_nodate}, such as operating a software development studio entirely through AI\cite{qian_communicative_2023}. Concurrently, LLM exhibit exceptional in-context learning\cite{noauthor_230100234_nodate} and few-shot learning capabilities\cite{brown_language_2020}, enabling them to continually learn during human interaction. This allows their collaborative abilities and responsibilities to grow and adapt dynamically\cite{noauthor_230102111_nodate}. Unlike previous Human-AI interaction patterns, where AI systems are often viewed as tools for processing fixed procedures, LLM can be seen as assistants or companions capable of continuous mutual learning and joint development\cite{zheng_synergizing_2023}. 

\subsubsection{Intention Understanding — Natural Language Based Interaction and Collaboration.}
Language and text serve as the primary carriers of information for human communication and collaboration\cite{noauthor_communication_nodate}. The capability of AI systems to comprehend human natural language significantly impacts the pattern and efficiency of the interaction process . Prior to the introduction of language models\cite{jing_survey_2019}, AI's capacity to process natural language was limited. Statistical models, such as the bag-of-words model\cite{zhang_understanding_2010}, and early deep learning techniques were usually just capable of language embedding within certain corpora, unable to achieve an understanding and generation of complex semantics.\\
users\indent With the development of self-supervised training\cite{noauthor_technologies_nodate}, the concept of pre-trained models has been widely adopted in the field of natural language processing (NLP)\cite{noauthor_pre-trained_nodate}. From the word2vec\cite{noauthor_using_nodate} to BERT\cite{devlin_bert_2019} and now the widely used GPT\cite{noauthor_190900512_nodate}, AI's capability to understand language has continuously improved. Presently, large language model, trained on massive corpora, exhibit exceptional intent recognition and natural language interaction abilities. They can respond to human needs based on natural language prompts\cite{cao_comprehensive_2023}, which significantly lowers the barrier for human to engage in collaboration with AI. Nowadays, a considerable number of interaction systems based on natural language have been proposed\cite{topsakal_creating_2023}. It is observable that interactive collaboration based on natural language prompts is gradually becoming a mainstream and promising mode of human-AI interaction\cite{noauthor_why_nodate}.
\subsubsection{Creativity — the ability to create new content.}
AIGC has received widespread attention in recent years, but it is not a novel application scenario for AI. As early as 1950, there were research focus on speech generation using the Hidden Markov Model (HMM)\cite{ide_hidden_1997}. However, due to the limitations in AI's generative capabilities, early generation tasks primarily focused on text-based content creation\cite{iqbal_survey_2022}. With the advancement of computer vision technologies, generative tasks have progressively expanded into the realm of imagery. Notable examples like the Generative Adversarial Networks (GANs)\cite{goodfellow_generative_2020} and the diffusion generative models\cite{noauthor_denoising_nodate}. At this stage of technology, AI is capable of performing various image adjustments, such as stylization and filtering.\\
\indent In recent years, with the growth of data and model sizes, large language model have exhibited exceptional generative capabilities. They can integrate user instructions to produce creative textual and visual content\cite{wu_ai_2021}, the quality and usability of which have significantly improved compared to traditional models\cite{zhang_complete_2023}. Against this backdrop, many conventional creative processes have begun to incorporate AI for collaborative production, such as scriptwriting, copywriting, graphic design, 3D modeling, and more.
\subsection{The Current Review of Human-AI Interaction}
\subsubsection{Different understanding of A in HAII Review.}
Human-AI Interaction is a frequently discussed topic within the Human-Computer Interaction field, primarily focusing on the interaction process between human and AI system. To date, there have been many comprehensive review studies on Human-AI Interaction, but the "A" (AI) of interest in these studies can vary considerably depending on the technological and research context. In this article\cite{noauthor_guidelines_nodate}, the researchers categorizes AI systems based on user visibility into AI-infused system and backend algorithms, then specifically targeting user-facing systems. Some researchers discuss the prospects for collaboration between human and AI with general capabilities\cite{noauthor_future_nodate}. Additionally, many studies define their focus through distinct application scenarios. For instance, several papers concentrate on the medical domain within article\cite{noauthor_ijerph_nodate}\cite{noauthor_jmir_nodate}\cite{noauthor_healthcare_nodate}, while others address the educational field\cite{noauthor_sustainability_nodate}\cite{bozkurt_unleashing_2023}, and yet another article\cite{patil_artificial_2023} is centered on management application scenarios.
\subsubsection{Current Mapping of HAII }
Numerous review studies have utilized diverse mapping approaches to systematically organize the extensive literature on Human-AI Interaction. In this study\cite{capel_what_2023}, the authors categorize the literature based on two key dimensions: the state of AI as either in design or usage, and the degree to which AI is human-centered. Conversely, this paper\cite{kim_one_2023} employs a clustering method and post-principal component analysis to categorize AI roles, focusing on human interaction and classifying AI into four distinct roles. The practice of defining and differentiating HAII research based on AI roles is prevalent in the HAII community. In this research\cite{sundar_rethinking_2022}, researchers defining four roles based on AI's communicative behavior. And in this work\cite{noauthor_bad_nodate} ,Researchers define four AI roles based on the level of moral conflict encountered in human-AI interactions.

\section{Method}
This review focuses on the interaction patterns between human and large language model. Accordingly, we have selected papers that explicitly identify with and contribute to the paradigms and patterns of Human-LLM Interaction. The papers in this systematic review have been gathered since the release of LLM. We primarily used the Association for Computing Machinery Digital Library (ACM DL) as our source for finding relevant articles on Human-LLM Interaction.\\
\indent Guided by this core concept, we conducted two rounds of literature screening, identifying 106 publications from an initial pool of 1398. We then employed a combination of manual scoring and machine clustering for mapping. In the mapping process, we adopted an approach similar to the one outlined in framework\cite{capel_what_2023}, determining the evaluation approach for Human-LLM interaction patterns through multiple workshop rounds. This included defining the map's perspectives, dimensions, concepts, and evaluation criteria. We applied these criteria to perform manual scoring and machine clustering of the literature. At last we defined the characteristics and differences of these clusters, which illustrated the current research distribution within this map.
\subsection{Search and Selection }
\begin{figure}[h]
    \centering
    \includegraphics[scale=0.12]{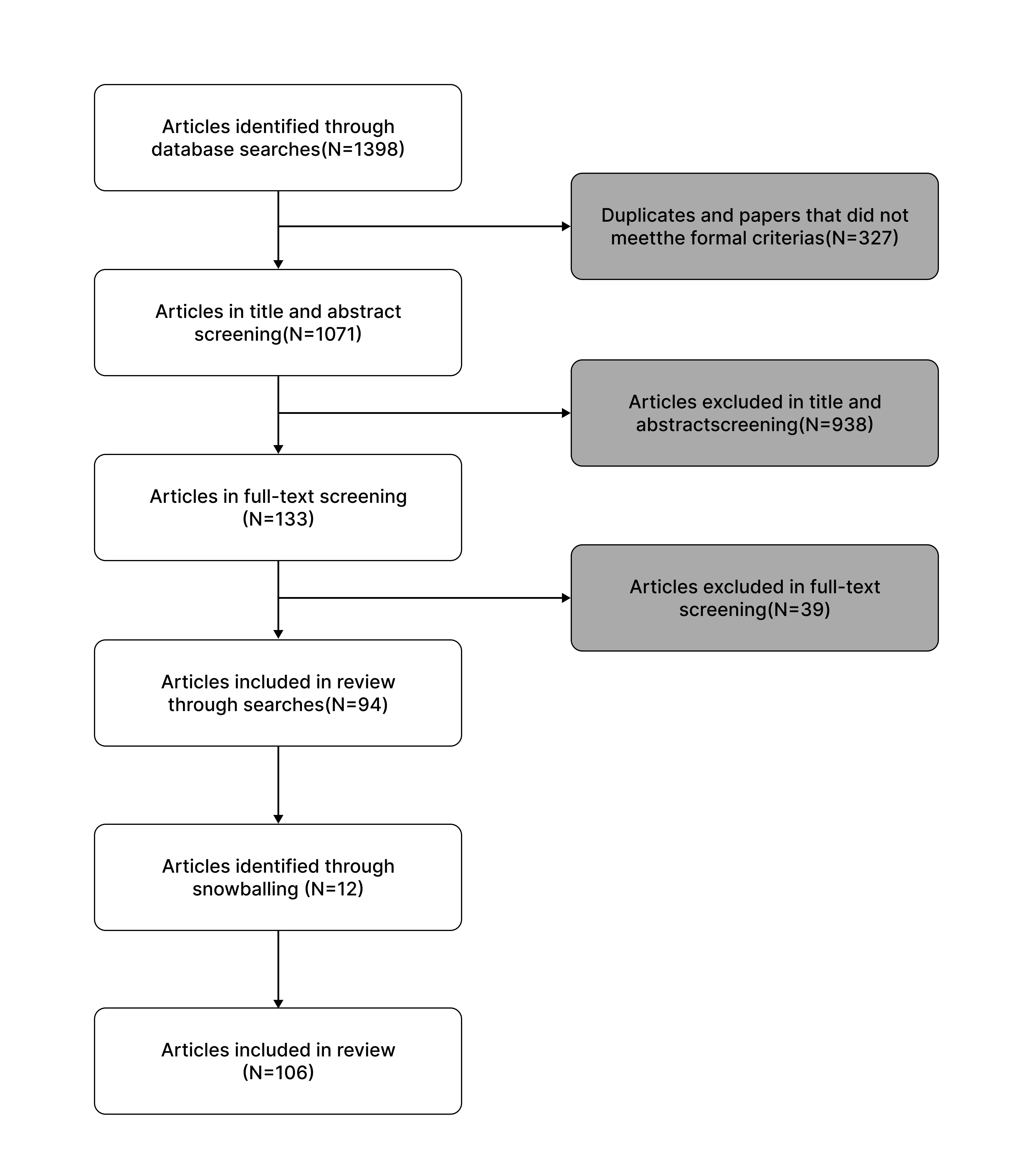}
    \caption{Flow diagram of the database searches and article screenings.} \label{fig1}
\end{figure}

In the literature search phase, we used the following keywords :
\\
( “large-language model” OR  “chatgpt” OR  “gpt” OR “LLM ” ) AND ( “human” OR  “user” OR  “interaction” OR  “human-centered ai” OR “co-creation ”  OR  “coordination” OR “ collaboration ”  OR  “user experience” OR “UX ” OR “interaction design ” OR “prompt design ”  OR  “user study” OR “user perception” OR “tool”). 
\\
\indent
This resulted in a total of 1398 initial items from the ACM Digital Library. We also seek for more information on the criteria or process used to significantly narrow down the selection. For example, "Following two rigorous rounds of manual review, prioritizing relevance to Human-LLM Interaction patterns, we refined our selection to 106 pertinent articles(see Fig.~\ref{fig1}).
\\
\indent In the first round of selection, our criteria were twofold: whether the article was published in international journals or major conference proceedings in the HCI field, and whether it focused on general interaction paradigms instead of specific application scenarios of LLM capabilities. 
\\
\indent The second round of selection centered on whether the studies addressed user perspectives in the interaction process with LLM, rather than solely focusing on LLM-related technical applications.To achieve these objectives, we established the following criteria:
\begin{enumerate}% 大写罗马数字作为列表标签
        \item The article must be based on systems that involve interaction between human and LLM (including applications layered on LLM).
        \item These articles must present at least one user study to ensure that the LLM-based system is directly user-facing.
        \item These articles should not focus solely on LLM applications within a specific domain, but rather on the general modes of interaction between LLM and human.
\end{enumerate}
%筛选流程图
\subsection{Mapping}
\begin{figure}
\centering
\includegraphics[width=\textwidth]{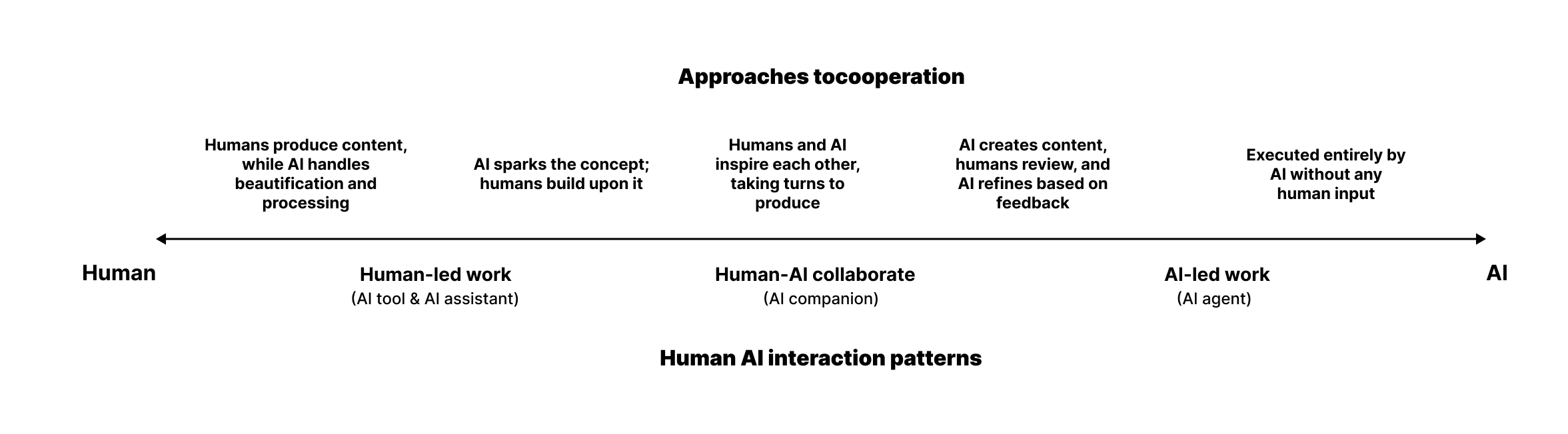}
\caption{Types of task division between human and AI in work.} \label{fig2}
\end{figure}

\begin{figure}
\centering
\includegraphics[width=\textwidth]{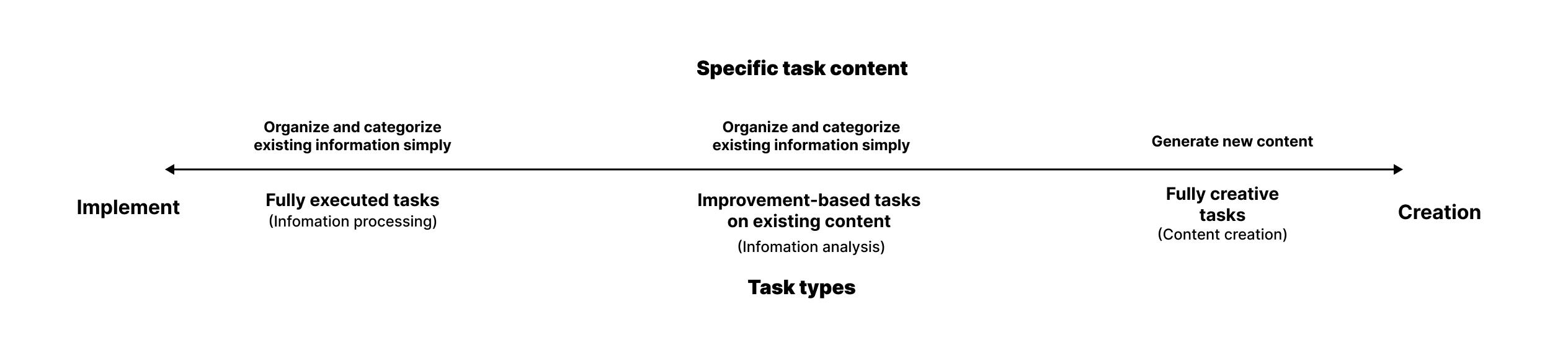}
\caption{Innovation level of content produced through human-AI task.} \label{fig3}
\end{figure}

Throughout the mapping process, we primarily referenced the workflow of mapping methods by\cite{capel_what_2023}. Considering that the existing mapping dimensions are not fully compatible, we need to summarize the article characteristics of articles included in review in order to propose new dimensions and mapping standards. 
\\
\indent During the review and analysis of selected articles, we observed that in collaborative tasks involving human and AI, the main factors influencing the interaction patterns are the collaborative relationship between human and AI, as well as creative level of tasks undertaken by AI. Therefore, we drew upon Ding, Zijian,Chan, Joel's methodology\cite{noauthor_mapping_nodate} in studying Human-AI Interaction patterns. Collaboration and creativity are being chosen as the main perspectives for our mapping. 
\\
\indent Building upon these two analytical perspectives\cite{capel_what_2023,noauthor_mapping_nodate}, we contemplated which specific measurement dimensions could be used. We analyzed the characteristics of human involvement-AI autonomy and complexity-creativity of 110 selected articles. After thorough discussions, we ultimately established two evaluative dimensions: Human-AI and Implement-Creation.
 \begin{enumerate}% 大写罗马数字作为列表标签
        \item Collaboration dimension corresponds to the Human-AI, reflecting the collaborative relationship and the decision-making dominance of human and AI. 
        \item Creativity dimension corresponds to Implement-Creation, measuring the type of tasks handled by AI in the collaboration process, to reflect AI’s creative contribution  in the system.
\end{enumerate}
\indent Upon determining these dimensions, we aimed to establish a set of mapping standards to position the articles on the corresponding coordinates. We defined the conceptual stages for each dimension after multiple rounds of discussion. In the Human-AI dimension, we categorized AI roles into four stages: tool, assistant, companion, agent, corresponding to four stages from human to AI decision-making. In the Implement-Creation dimension, we summarized three concepts: process, analysis, create. These conceptual stages aid in understanding the types of tasks handled by AI under the entire collaboration goal, facilitating more precise article classification and scoring criteria.
\\
\indent Next, based on these conceptual stages, we established criteria for each stage to more accurately map the articles onto the map.
 \begin{enumerate}% 大写罗马数字作为列表标签
        \item In the Human-AI dimension, we defined five mark points(see Fig.~\ref{fig2}), -2: Human produce content, while AI handles beautification and processing; -1:AI sparks the concept then human execute upon it;0: human and AI inspire each other, taking turns to produce;1: AI creates content, then human review, and AI refines based on feedback;2: Executed entirely by AI without any human input. These levels evaluate the role and decision-making relationship of AI under the entire collaboration goal. To precisely evaluate articles at the same point, we refined the scoring to one decimal place. For instance, for articles with values -2, we located their position in the X range [-2.5, -1.5].
        \item In the Implement-Creativity dimension, we defined three mark points(see Fig.~\ref{fig3}), Organize and categorize existing information simply, 1:Organize and categorize given data; 2:Analysis and form opinions based on the given data; Generate new content based on given data. These dimensions evaluate the creativity of the task AI undertakes. By refining the scoring to one decimal place, we enhanced the precision of our evaluations, allowing for nuanced distinctions between articles that scored similarly. For instance, for articles with values -1, we located their position in the Y range [-1.5, -0.5].
\end{enumerate}

\begin{figure}[htbp]
    \centering
    \begin{minipage}[t]{0.35\linewidth} % 左边图片的容器宽度调整为40%
        \centering
        \includegraphics[width=\linewidth]{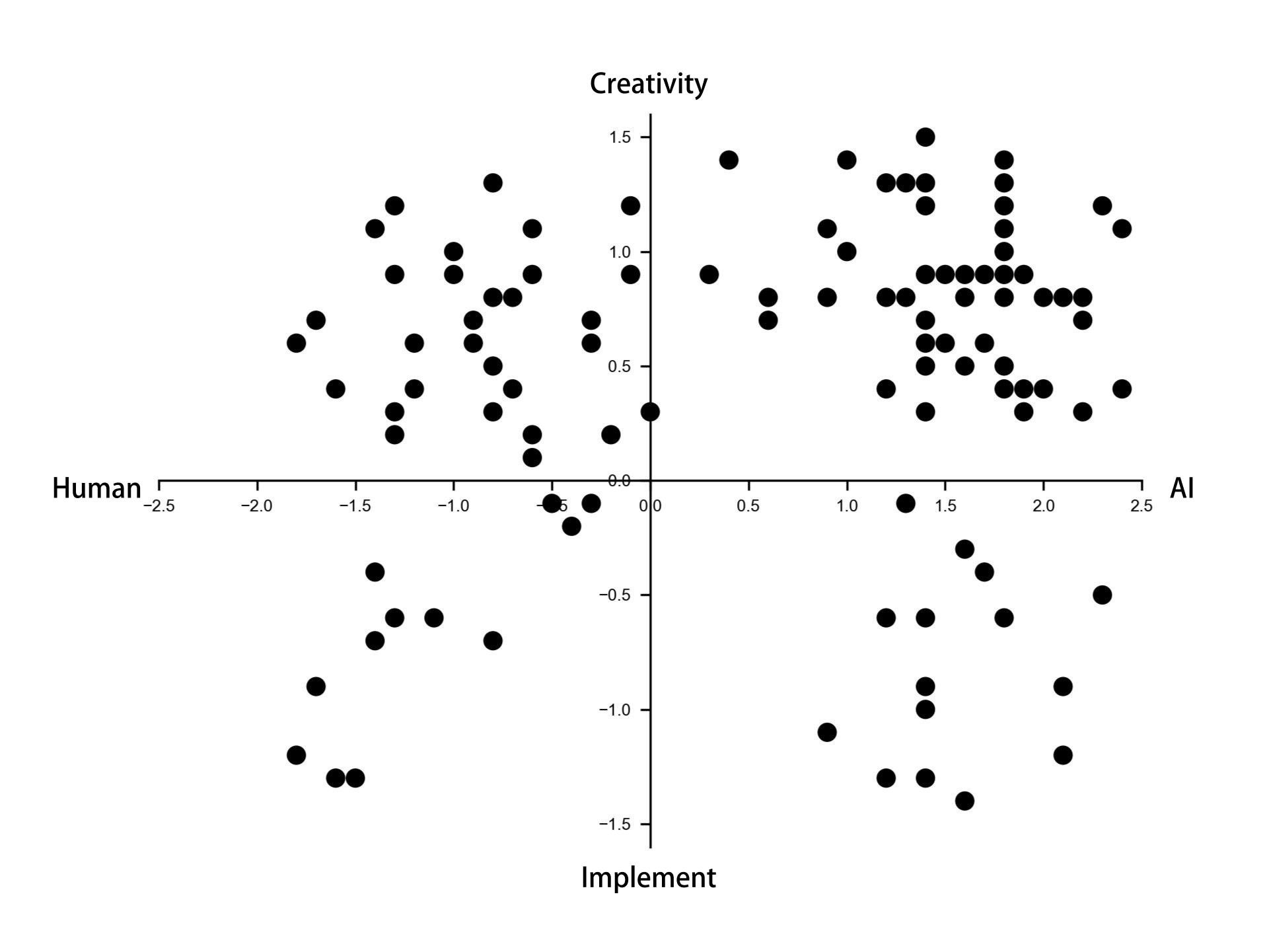}
        \caption{The overall scores of the article in the map.}
        \label{fig:map}
    \end{minipage}
    \hfill
    \begin{minipage}[t]{0.55\linewidth} % 右边图片的容器宽度调整为50%
        \centering
        \includegraphics[width=\linewidth]{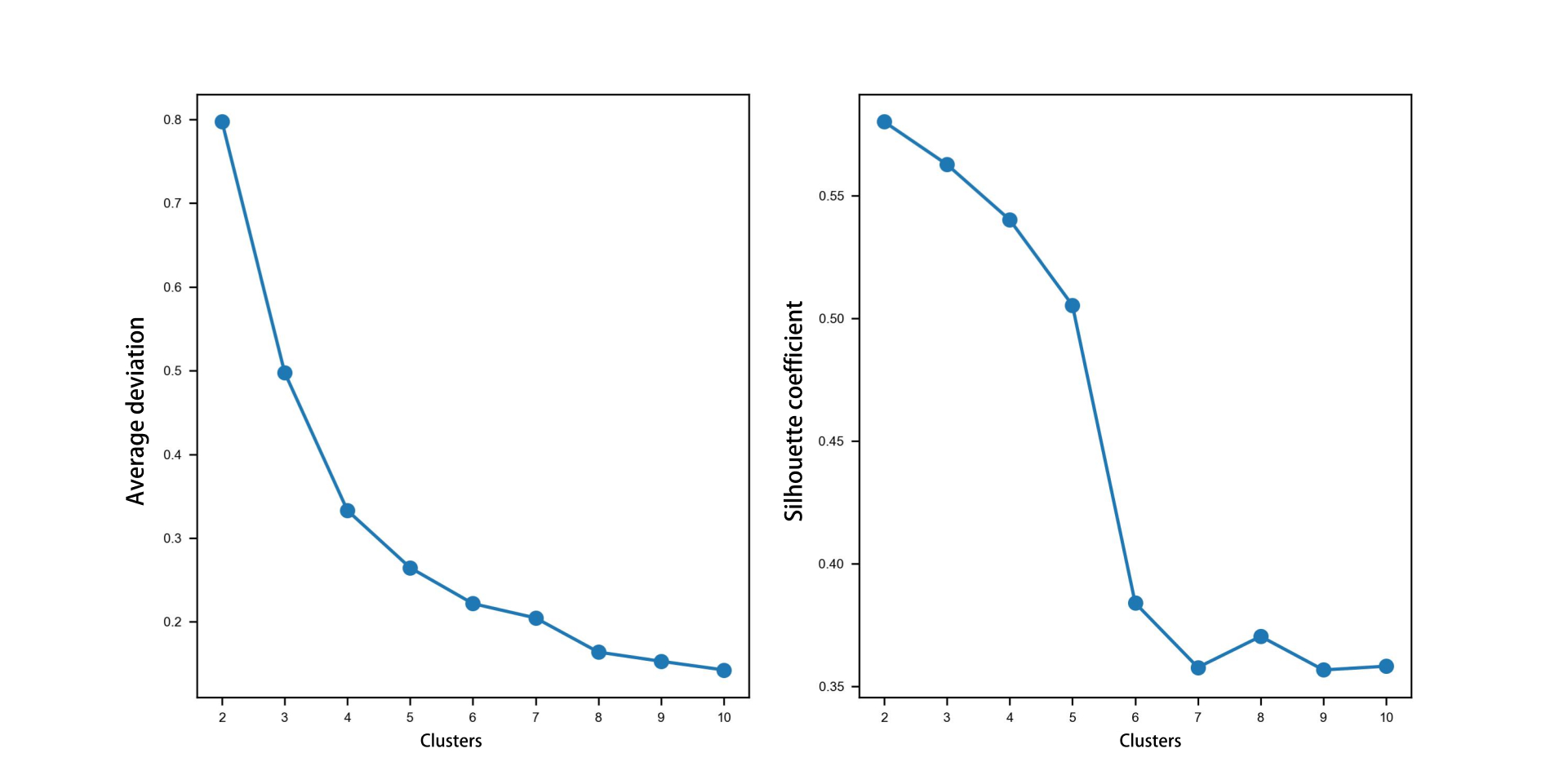}
        \caption{Elbow (left) and silhouette (right) method for determining the optimal number of clusters.}
        \label{fig5}
    \end{minipage}
\end{figure}

\indent We conducted three rounds of scoring for the 110 publications, with each round designed to incrementally refine and validate the objectivity of each article's score.The overall scores of the article(see Fig.~\ref{fig:map}).
\\
\indent We then used the K-means algorithm to cluster the 110 scored articles. We chose the k parameter as 4.This decision was based on the average deviation, which measures variability among the data, and the silhouette coefficient, an indicator of how similar an object is to its cluster compared to other clusters(see Fig.~\ref{fig5}).
\section{Result}
Based on our clustering, we defined four primary interaction pattern classifications between human and LLM: 1) Processing Tool, where LLM perform specific, directed tasks; 2) Analysis Assistant, providing analytical support to human input; 3) Processing Agent, engaging in more autonomous tasks with human; 4) Creative Companion, where LLM contribute creatively and collaborative in tasks.
\begin{figure}[h]
    \centering
    \includegraphics[scale=0.18]{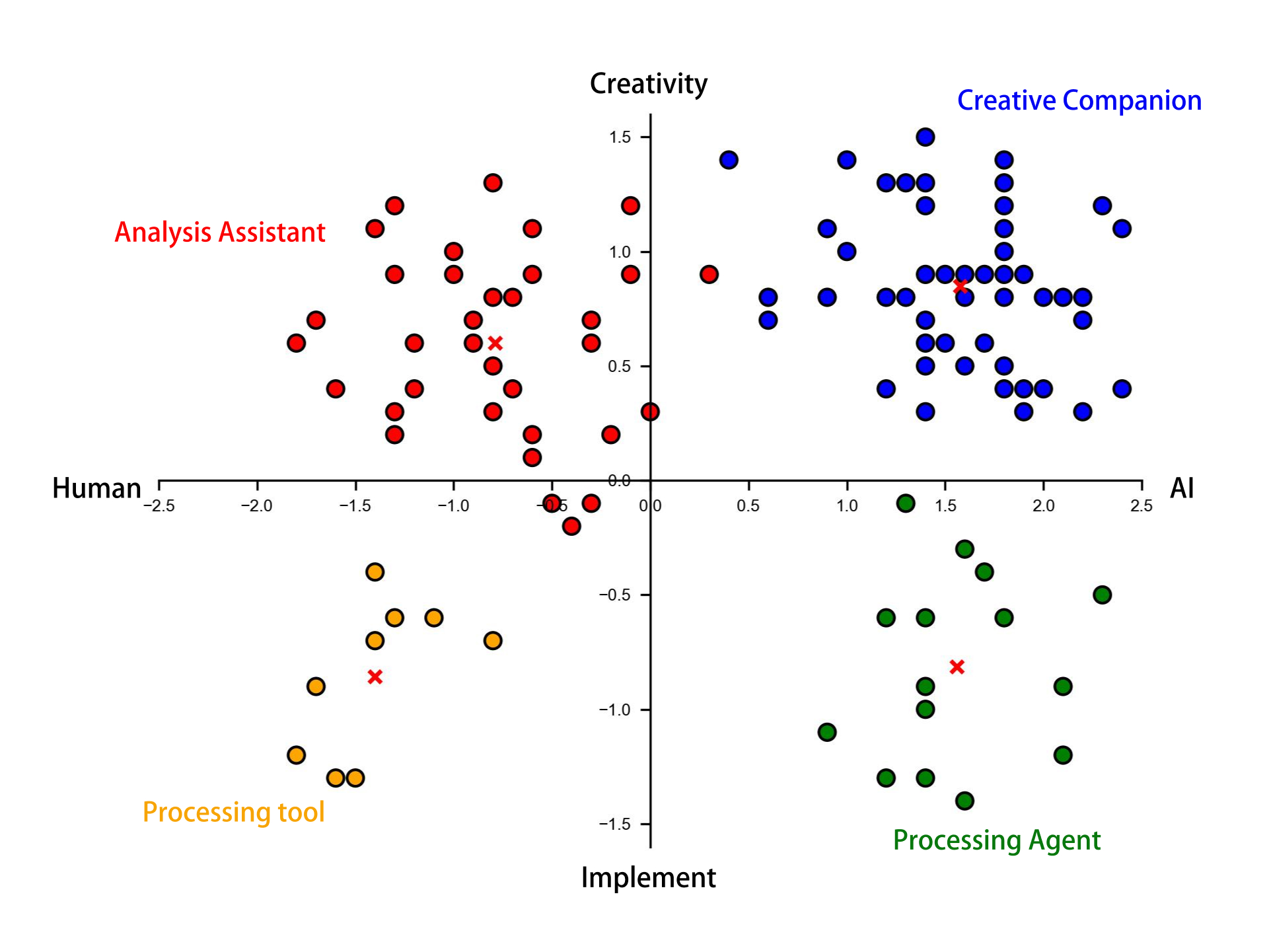}
    \caption{Map of the feld of human-LLM interaction patterns. The chart displays the comparative volume of research papers in each sector.} 
    \label{fig6}
\end{figure}
\subsection{Processing tool}
The articles labeled orange\cite{penha_improving_2023,lu_readingquizmaker_2023,lee_promptiverse_2022,mcnutt_design_2023,karinshak_working_2023,yang_harnessing_2023,wang_enabling_2023}(see
Fig.~\ref{fig6}), have the following characteristics. Regarding creativity, AI's role is confined to processing given data, lacking original creative input. It is usually responsible for static and specific tasks, processing data and producing outputs in a predetermined format, such as sequence-to-sequence, classification, recognition. From a collaborative standpoint, the work executed by AI and the result output mainly serve as raw material for human's decision-making or used as input for other modules. AI seldom takes the responsibility for decision-making while collaborating with human, it only provides inductions and summaries derived from the processing of existing data. In this scenario, human predominantly make decisions and form opinions, utilizing AI's work as a foundational basis. Therefore, we categorize the AI in this cluster as a Processing tool.
\\
\indent In this work\cite{wang_enabling_2023}, researchers employ large language model to facilitate user conversational interactions with smartphone interfaces. In terms of collaboration, LLM serve as tools for operation translation from natural language, while in creative aspects, the AI merely performs the function of mapping input data to a limited space of user interface actions. In study\cite{penha_improving_2023}, the application of LLM focuses on traditional and task-specific functions like information retrieval, highlighting their potential in this domain. In study\cite{yang_harnessing_2023}, AI-assisted treatment plans, while generated based on patient data, necessitate further evaluation and refinement by human physicians.
\subsection{Analysis assistant}
The articles labeled red \cite{singh_where_2022,bhat_interacting_2023,liu_will_2022,shakeri_saga_2021,yuan_wordcraft_2022,gero_social_2023,wu_ai_2022,fu_comparing_2023,kim_metaphorian_2023,buschek_impact_2021,karolus_your_2023,goodman_lampost_2022,biermann_tool_2022,jakesch_co-writing_2023,bhavya_cam_2023,robe_designing_2022,ross_programmers_2023,robe_pair_2022-1,jonsson_cracking_2022,weisz_better_2022,al_madi_how_2023,jiang_discovering_2022,kazemitabaar_studying_2023,barke_grounded_2023,liu_what_2023,danry_dont_2023,gero_sparks_2022,wang_designing_2023,wallace_embodying_2023,robertson_wait_2021,guo_towards_2023,zamfirescu-pereira_why_2023,fraile_navarro_collaboration_2023,ruoff_onyx_2023,huang_scones_2020,van_der_burg_objective_2023}(see
Fig.~\ref{fig6}), exhibit the following characteristics. From a collaborative perspective, the task of AI is not confined to summarizing data and returning processing result, rather, it involves formulating opinions based on provided information. AI's opinion would influence human, thereby aiding in collaboration decision-making. From a task creativity standpoint, AI-generated outcomes involve summarizing different perspectives and creatively interpreting the input data. There is a clear hierarchical and structural difference between AI system's input and output, the representation of input data in the output is highly abstract. Consequently, the work performed by AI exhibits a significant degree of creativity. In this cluster, AI functions as an Analysis assistant supporting human decision-making in collaboration, rather than merely serving as a tool.
\\
\indent In these studies\cite{singh_where_2022,bhat_interacting_2023,liu_will_2022,shakeri_saga_2021,yuan_wordcraft_2022,gero_social_2023,wu_ai_2022,fu_comparing_2023,kim_metaphorian_2023,buschek_impact_2021,karolus_your_2023,goodman_lampost_2022,biermann_tool_2022,jakesch_co-writing_2023,bhavya_cam_2023}, researchers discuss how human collaborate with AI in writing tasks, such as replying to emails or creating stories, AI would continously provide the reference or suggestion to facilitate human's writting. The following research\cite{robe_designing_2022,ross_programmers_2023,robe_pair_2022-1,jonsson_cracking_2022,weisz_better_2022,al_madi_how_2023,jiang_discovering_2022,kazemitabaar_studying_2023,barke_grounded_2023,liu_what_2023} focuses on the interactive programming system that enables continuous mutual suggestions and pairs programming with human for programming tasks. Studies\cite{danry_dont_2023,gero_sparks_2022,wang_designing_2023,wallace_embodying_2023,robertson_wait_2021,guo_towards_2023,zamfirescu-pereira_why_2023,fraile_navarro_collaboration_2023,ruoff_onyx_2023} delve into human reflection on AI-generated content and its capacity for inspiration, often focusing on Human-AI Interaction aspects like explainability and transparency. And these two studies\cite{huang_scones_2020,van_der_burg_objective_2023} in question are centered on the application scenarios of AI-assisted image creation, where individuals draw inspiration from designs generated by AI, and through providing feedback, collaborate with AI in the process of image creation.
\subsection{Creative companion}
The articles labeled blue \cite{cho_areca_2023,rajcic_message_2023,srivastava_response-act_2023,xiao_inform_2023,arakawa_catalyst_2023,nguyen_extracting_2023,xu_generating_2023,pang_empmff_2023,xu_cosplay_2022,zhao_multiple_2020,pataranutaporn_living_2023,si_why_2022,ye_reflecting_2022,jo_understanding_2023,zamfirescu-pereira_herding_2023,xygkou_conversation_2023,scott_you_2023,yin_ctrlstruct_2023,niwa_investigating_2023,he_unified_2022,sharma_towards_2021,elgarf_creativebot_2022,hada_rexplug_2021,noauthor_social_nodate,cao_multi-modal_2022,liu_ssap_2022,ashby_personalized_2023,laban_newspod_2022,khanmohammadi_prose2poem_2023,suh_codetoon_2022,jones_embodying_2023,rajcic_mirror_2020,liu_opal_2022,huynh_argh_2021,lee_dapie_2023,lee_coauthor_2022,chung_talebrush_2022,yanardag_shelley_2021,chiou_designing_2023,liu_3dall-e_2023,brie_evaluating_2023,lanzi_chatgpt_2023,noauthor_co-writing_nodate,dang_choice_2023,tholander_design_2023,chen_marvista_2023,metzler_rethinking_2021}(see Fig.~\ref{fig6}), exhibit the following characteristics. From a collaborative perspective, AI is expected to handle more complex tasks and take on greater responsibility in decision-making. This primarily manifests in AI handling high-level tasks that require a combination of various specific skills working in concert. AI needs to independently decide when to utilize a particular skill at the right time. This also allows AI to reduce the dependency on human during task execution, granting it higher independence and autonomous decision-making authority. On the other hand, from a creative standpoint, human convey their ideas directly to large language model, which autonomously extend and create based on these inputs. AI will generate creations or answers based on the human's description, autonomously extending and interpreting these inputs. 
 The content generated by AI could be inspirational, exhibiting elements of design and art. 
\\
\indent In studies\cite{cho_areca_2023,rajcic_message_2023}, researchers explore AI's application in furniture, endowing it with autonomy and creativity to perceive events and generate content like diaries\cite{cho_areca_2023} and work suggestions\cite{arakawa_catalyst_2023}. In studies\cite{srivastava_response-act_2023,xiao_inform_2023,arakawa_catalyst_2023,nguyen_extracting_2023,xu_generating_2023,pang_empmff_2023,xu_cosplay_2022,zhao_multiple_2020,pataranutaporn_living_2023,si_why_2022,ye_reflecting_2022,jo_understanding_2023,zamfirescu-pereira_herding_2023,xygkou_conversation_2023,scott_you_2023,yin_ctrlstruct_2023,niwa_investigating_2023,he_unified_2022,sharma_towards_2021,elgarf_creativebot_2022,hada_rexplug_2021,noauthor_social_nodate,cao_multi-modal_2022,liu_ssap_2022}, they primarily presents chatbot that utilize natural language communication as their form of interaction to assist in responding to users' open-ended queries, such as psychological healing\cite{srivastava_response-act_2023} and history learning\cite{pataranutaporn_living_2023}. In studies\cite{ashby_personalized_2023,laban_newspod_2022,khanmohammadi_prose2poem_2023,suh_codetoon_2022,jones_embodying_2023,rajcic_mirror_2020,liu_opal_2022,huynh_argh_2021,lee_dapie_2023,lee_coauthor_2022,chung_talebrush_2022,yanardag_shelley_2021,chiou_designing_2023,liu_3dall-e_2023,brie_evaluating_2023}, AI systems are tasked with transforming input data into stylized and creative outputs, such as converting prose into Persian poetry\cite{chiou_designing_2023}, or generating comics from programming code\cite{suh_codetoon_2022}. The research in group\cite{lanzi_chatgpt_2023,noauthor_co-writing_nodate,dang_choice_2023,tholander_design_2023,chen_marvista_2023,metzler_rethinking_2021} focuses on AI refining and elaborating on creative content from rough descriptions, such as generating complete game designs\cite{lanzi_chatgpt_2023} or distilling literary ideas from existing articles\cite{metzler_rethinking_2021}.
\subsection{Processing agent}
The articles labeled green\cite{wang_documentation_2022,alonso_del_barrio_framing_2023,petridis_anglekindling_2023,august_paper_2023,sun_investigating_2022,zhou_story_2019,shen_kwickchat_2022,li_personalized_2023,xue_prefrec_2023,liao_designerly_2023,zhang_concepteva_2023,wang_reprompt_2023,hamalainen_evaluating_2023,sekulic_evaluating_2022,le_socialbots_2022}(see
Fig.~\ref{fig6}), exhibit the following characteristics. In the process of collaboration with human, AI will carry out a series of complex tasks, leveraging its general capabilities for decision-making and task processing. The system exhibits traits akin to a Creative Companion, implying its ability to autonomously execute complex tasks and make decisions during key collaborative phases. From a creative perspective, these tasks generally pertain to the organization and descriptive summarization of data. The AI's output, while only a summary of the input data, shows a higher complexity level than a standard processing tool, though it does not exhibit additional creativity.
\\
\indent In this series of studies\cite{wang_documentation_2022,alonso_del_barrio_framing_2023,petridis_anglekindling_2023,august_paper_2023,sun_investigating_2022,zhou_story_2019,shen_kwickchat_2022,li_personalized_2023,xue_prefrec_2023,liao_designerly_2023,zhang_concepteva_2023,wang_reprompt_2023}, AI operates with a high degree of autonomy in analyzing relevant information and summarizing content. This includes tasks such as generating descriptive documents that outline human work\cite{wang_documentation_2022}, selecting story endings from existing information\cite{zhou_story_2019}, or presenting medical papers interactively\cite{august_paper_2023}. This research group utilizes AI to simulate user operations, including emulating user responses in completing questionnaires\cite{hamalainen_evaluating_2023} and mimicking user behavior within a search system\cite{sekulic_evaluating_2022}.
\section{Discussion}
\subsection{Mapping Methodology Based on Human and Algorithmic Approaches}
This study uniquely contributes to both theoretical understanding and practical applications in mapping Human-LLM Interaction patterns. In terms of mapping methods, the scarcity of comprehensive review studies on Human-AI Interaction, particularly in the domain of interaction with LLM, results in a lack of directly applicable mapping approaches in this field. Therefore, we developed mapping methods specifically designed to capture the difference and complexities of human-AI interaction patterns. Our method consists of five steps: defining the perspective, identifying dimensions, associating relevant concepts within dimensions, establishing evaluation criteria based on concepts, scoring and clustering
\\
\indent Firstly, our observations from the selected articles revealed two key influences of LLM on HAII patterns: the relationship of Human-AI Collaboration and their distinct task responsibilities. Hence, we chose collaboration and creativity as the research perspectives for this review. Secondly, we identified studies on evaluating collaboration and creativity to determine the two evaluative dimensions: Human-AI and Implement-Creation. Thirdly, upon the classification concepts we found, combined with the characteristics of the selected articles and after multiple rounds of discussion, we finalized the conceptual stages for each dimension. This process aids in a more intuitive understanding of different task states at various stages and prepares specific standards for the two dimensions. Next, through discussion and based on the different conceptual stages, we established criteria for each stage, enabling us to more accurately map articles to their corresponding positions on the map. Finally, based on these scoring criteria, we conducted multiple rounds of scoring for the 110 selected articles and used the K-means algorithm for machine clustering, presenting an objective overview of the main type distributions in current research.
\\
\indent The conceptual stages we selected are partly similar to those defined in the source literature but are not completely identical. For instance, the definitions of AI tools and AI assistants, according to the former definition, AI tools are characterized as having low human involvement and low AI autonomy, and AI assistants are represented as having low human involvement and high AI autonomy. This aligns with our positioning of AI tools and AI assistants in the Human-AI dimension. While as we also consider the type of tasks each concept should handle in our definitions, AI tools in our study are involved in low-level tasks, whereas AI assistants handle the creation of viewpoints, which are relatively higher-level tasks. Our need to categorize conceptual stages into two distinct dimensions results in variations from the positions outlined in the referenced literature.
\subsection{Differences between clusters}
Building on our mapping and clustering results, we delved into a detailed analysis, focusing on the quantitative distribution and distinct characteristics of each cluster.
\subsubsection{Research Interests}
\begin{figure}[h]
    \centering
    \includegraphics[scale=0.15]{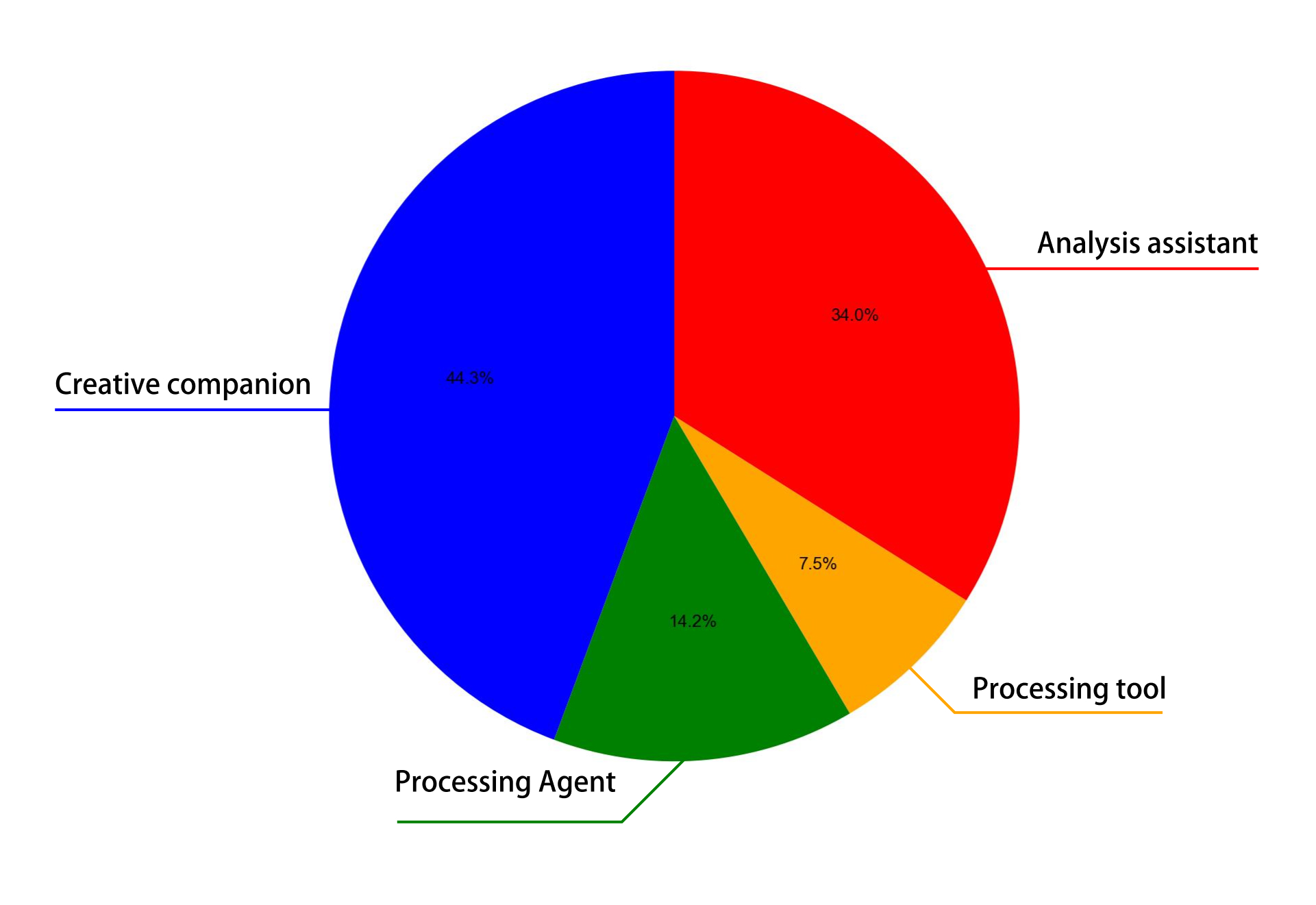}
    \caption{The current research interests in the field of Human-LLM Interaction patterns.} \label{fig7}
\end{figure}

We analyzed the number of papers in each cluster and found that over three-quarters of the articles focused on AI’s involvement in creative tasks(see Fig.~\ref{fig7}).
\\
\indent Almost half of the papers were concentrated in the Creative companion cluster. This indicates that the AI systems based on large language model exhibit a heightened state of creativity and autonomy in task execution. This aligns with the current intuitive understanding that the performance enhancements of LLM compared to traditional AI are primarily evident in creativity and the execution of complex tasks. Therefore, more research interests are focusing on creative content generation, exploring how to grant AI more autonomy in decision-making within collaborative work with human.  
\\
\indent Simultaneously, the tasks executed by AI are becoming more creative. This conclusion can be drawn from the observation that over three-quarters of the articles are concentrated in the clusters of Creative Companion and Analysis Assistant, both of which are positioned high on the creative dimension map. Compared to traditional AI tasks (such as classification, prediction, identification of given data), AI is now performing more creative tasks like image generation\cite{liu_3dall-e_2023}, scriptwriting\cite{yuan_wordcraft_2022}. AI, while operating within the framework of human guidance and requirements, goes beyond mere interpretation, creatively generating new content and insights."
\subsubsection{Category Differences}
To clarify the differences between similar or easily confused categories, we engaged in discussions to more precisely define their characteristics.
\\
\indent From the perspective of Collaboration, the differences between clusters cannot be simply summarized as the specific tasks AI performs. Rather, they are the relative relationship of these tasks to the overall collaborative goal, specifically manifesting in the extent to which AI’s work influences decision-making in the collaborative process. Consider the scriptwriting task, for instance: in some studies\cite{shakeri_saga_2021}, AI-generated scripts offer inspiration that human evaluate and potentially adopt in their writing. In other tasks, AI first collects human ideas or pre-work,and then generates the final script content\cite{lee_coauthor_2022}. Hence, in the latter case, the positioning on Collaboration mapping is closer to Human. These differences are mainly evident when comparing Creative Companion with Analysis Assistant. 
\\
\indent We also find differences in the Collaboration dimension reflected in the complexity of tasks AI performs, such as AI's responsibilities in providing optimization suggestions to the system. Some works focus on using large language model to do the simple task like operation signal transformation\cite{wang_enabling_2023}. Other works extend this technology to simulate and deduce a complex system behavior\cite{sekulic_evaluating_2022}, where the latter handles more complex tasks and requires AI to exercise more autonomous decision-making and capacity planning, thus being mapped closer to AI on the Collaboration dimension. This difference is primarily manifested between Processing Tool and Processing Agent.
\\
\indent From a Creativity perspective, differences between categories stem not merely from the AI-generated content's structure or type but from the degree of creativity relative to the input. It is more about the degree of creativity in AI's output relative to the input it receives. For example, in code generation tasks, some works generate code based on detailed functional descriptions by professionals, translating from structural description like information in jupyter notebook to code\cite{mcnutt_design_2023}. Others generate code around human needs and abstract descriptions\cite{jonsson_cracking_2022}. Compared to the former, the latter has more abstract input information, and the disparity between input and output data is more pronounced, thus being mapped closer to the Create in the Creativity dimension. This difference is evident between Analysis Assistant and Processing Tool. 
\\
\indent The same difference can also be seen in different chatbots. In chatbots engaged in specific interactions with human, some are not restricted in their reference to external knowledge or context, conducting conversations on any subject with human\cite{yin_ctrlstruct_2023}. In contrast, some chatbots are limited to explaining and describing given knowledge or documents\cite{petridis_anglekindling_2023}. Compared to the latter, the former produces content with higher degrees of freedom and is thus mapped closer to the Create position in the Creativity dimension. This difference is exemplified between Creative Companion and Processing Agent.
\subsection{About the vacancy in the mapping}
In our analysis, we identified a opbivous vacant space in the mapping(see
Fig.~\ref{fig6}). There are very few papers distributed within this area. Our explanation of this phenomenon is that the characteristics corresponding to the collaboration dimension in this interval(-1,1) require AI to provide continuous suggestions and participate in joint decision-making. Consequently, AI must focus on explaining its recommendations, prioritizing the transparency and interpretability of information. This necessitates that AI possesses a certain level of creative ability, such as generating explanations for its own work[148]. The task of providing explanations and clarifications inherently involves creativity, which accounts for the generally higher levels of creativity observed in AI personas within this specific interval.
\subsection{Future Directions}
As the understanding capabilities of large language model continue to improve, allowing for deeper comprehension of human expressions, the future will likely see LLM as companions in a wider array of specific application domains and collaborative tasks. Moreover, with the ongoing enhancement of LLM' generative abilities, their versatility, and transfer capabilities, the range of tasks in Human-LLM Interactions is set to expand, enabling LLM to undertake more creative work across various fields.
In practical use, future interactions with large language model are expected to continue exhibiting the current three trends of task division:
\begin{enumerate}% 大写罗马数字作为列表标签
        \item Human-led tasks: Here, LLM primarily handle process execution and some decisions that are less critical to the overall task. human take charge of core decision-making and control the overall direction of the task.
        \item Collaborative tasks: Both parties provide suggestions and jointly make decisions and execute the task.
        \item AI-led tasks: Human mainly provide input, while the overall task is decided and executed by the LLM.
\end{enumerate}
\section{Limitation and Future Work}
This article has certain limitations. During the literature search phase, we only organized articles published before July 2023. Given the rapid iteration of large language model related technologies, it is possible that new collaborative methods have emerged that were not included in this review. In the manual scoring phase, although the authors conducted three joint scoring sessions to be as objective as possible, manual analysis still retains a degree of subjectivity. To address this issue, future work could involve better integration of algorithms into the classification dimensions (such as using principal component analysis methods) to ensure a more objective assessment.
\section{Conclusion}
Interactions between human and AI are becoming increasingly common in today's society. This article organizes research related to Human-LLM Interaction modalities that have emerged since the advent of LLM. Adopting collaboration and creativity as research perspectives, it analyzes these articles from the Human-AI and Implement-Creation dimensions, constructing an overview and map of human-AI interaction research (HAII Research Overview and Map). On this basis, the study identified four main areas through machine clustering: Processing Tool, Analysis Assistant, Creative Companion, and Processing Agent. By conducting a secondary reading of articles within each cluster, we extracted the characteristics and research interests of each category and clarified the differences between them through discussion.
\\
\indent Overall, our review offers new perspectives and methods for understanding and evaluating Human-LLM Interaction patterns. It organizes existing research in the field based on this perspective, further classifies these studies through mapping, and clarifies the current status and challenges of each category.
\section{Acknowledgments}
This article received funding support from the Tongji University Graduate Important International Academic Conference Award Fund and the research project on key technologies of hydrogen energy town fuel cell integrated energy systems.

%\bibliographystyle{splncs04}
\bibliographystyle{unsrt}
\bibliography{lastused.bib}
\end{document}